\documentclass[12pt]{article}
\title{On compatibility of Bohmian mechanics with 
standard quantum mechanics}
\author{Hrvoje Nikoli\'c \\
Theoretical Physics Division, Rudjer Bo\v{s}kovi\'{c} Institute, \\
P.O.B. 180, HR-10002 Zagreb, Croatia \\
{\normalsize E-mail: hrvoje@thphys.irb.hr} \\
\makebox[1in]{} \\
}
\begin{document}
\maketitle
\begin{abstract}
It is shown that the apparent incompatibility of Bohmian mechanics with  
standard quantum mechanics, found by Akhavan and Golshani 
[quant-ph/0305020],
is an artefact of the fact that the initial wavefunction they 
use, being proportional to a $\delta$-function, is not a regular
wavefunction.  
\end{abstract}
PACS numbers: 03.65.Ta
\vspace{0.5cm}

\noindent

In \cite{ga}, Golshani and Akhavan argued that Bohmian mechanics
\cite{bohm} was not compatible with standard quantum mechanics. 
They studied a case in which a 2-particle wavefunction 
$\psi(x_1,y_1,x_2,y_2,t)$ describing a pair of 
entangled particles moving in the $x$-$y$ plane was such that the Bohmian 
trajectories of the particles obey
\begin{equation}\label{e1}
y_1(t)+y_2(t)=0,
\end{equation}
provided that the initial positions of the particles obey
\begin{equation}\label{e2}
y_1(0)+y_2(0)=0.
\end{equation}  
They have shown that (\ref{e1}) is incompatible with standard quantum 
mechanics and suggested that this could be used to distinguish 
experimentally between Bohmian mechanics and standard quantum mechanics.
Not surprisingly, an experiment \cite{brida} has confirmed  
standard quantum mechanics.

Struyve {\it et al.} \cite{str} criticized the conclusion
of Golshani and Akhavan. They stressed that, owing to the 
quantum equilibrium hypothesis \cite{durr}, it is not possible 
to achieve the initial condition (\ref{e2}) in a laboratory. 
Instead, initial positions are distributed according to the 
quantum mechanical role
\begin{equation}\label{e3}
\rho(x_1,y_1,x_2,y_2,0)=
\psi^*(x_1,y_1,x_2,y_2,0)\psi(x_1,y_1,x_2,y_2,0).
\end{equation}
The initial distribution (\ref{e3}) implies that particle 
positions are distributed according to the 
quantum mechanical role $\rho=\psi^*\psi$ at any time $t$. 
In this way, they have shown that there is no incompatibility 
between Bohmian mechanics and standard quantum mechanics, as far as 
the situation in \cite{ga} is studied. 

Recently, Akhavan and Golshani replied to the 
criticism of Struyve {\it et al.} by 
arguing \cite{ag} that the 
initial condition (\ref{e2}) is achieved in a situation in which 
the initial wavefunction $\psi(x_1,y_1,x_2,y_2,0)$ is 
proportional to 
\begin{equation}\label{e4}
\psi(y_1,y_2)=\delta(y_1-y_2),
\end{equation}
which, in a similar way, leads to (\ref{e1}) and thus to 
incompatibility
between Bohmian mechanics and standard quantum mechanics.
In this short paper we show that this result 
is an artefact of the fact that the wavefunction 
(\ref{e4}) is not a regular wavefunction. When 
the $\delta$-function in (\ref{e4}) is regularized in an 
appropriate way, no incompatibility
between Bohmian mechanics and standard quantum mechanics 
sustains.  

For simplicity, we suppress the $x$-dependence of the wavefunction. 
Since the wavefunction $\psi(y_1,y_2,t)$ satisfies
the Schr\"odinger equation, the density 
$\rho(y_1,y_2,t)=\psi^*(y_1,y_2,t)\psi(y_1,y_2,t)$ satisfies
the continuity equation
\begin{equation}\label{e5}
\frac{\partial\rho}{\partial t}+
\frac{\partial(\rho v_{y_1})}{\partial y_1}+
\frac{\partial(\rho v_{y_2})}{\partial y_2}=0,
\end{equation}
where $v_{y_k}(y_1,y_2,t)$ (with $k=1,2$) are the Bohmian velocities 
determined by the wavefunction $\psi(y_1,y_2,t)$. 
The fact that the assumption that particles are initially 
distributed according to $\rho(y_1,y_2,0)$ leads to the result 
that they are distributed according to $\rho(y_1,y_2,t)$ at any 
time $t$ is a direct consequence of Eq. (\ref{e5}). On the other hand, 
the result of Akhavan and Golshani \cite{ag} corresponds to a 
situation in which the initial distribution of particles 
(\ref{e2}) is given by $\rho(y_1,y_2,0)$, but at later times 
the distribution of particles is not given by $\rho(y_1,y_2,t)$. 
Therefore, their result must be a consequence of a failure in Eq. 
(\ref{e5}). 

It is clear that the failure is due to the use of the wavefunction
(\ref{e4}) proportional to a $\delta$-function. In fact, the 
$\delta$-function is not a function at all, but rather a 
functional, meaningful only when integrated over the argument 
of the functional. Moreover, the square of the $\delta$-function, 
which appears in $\rho$, is not well defined even as a 
functional. Therefore, Eq. (\ref{e5}) does not make sense 
for the wavefunction proportional to (\ref{e4}). In order 
to make sense of it, one has to regularize the irregular 
``function" $\delta(y)$ (where $y\equiv y_1-y_2$). The most natural 
regularization is to use a Gaussian with a 
small but finite width $\Delta y$
and then, at the end of calculation, 
to consider the limit $\Delta y\rightarrow 0$. 
With finite $\Delta y$, the initial 
positions of the particles do not satisfy (\ref{e2}). Instead, 
they are distributed according to the Gaussian with the width 
$\Delta y$. It is clear that for {\em any} 
finite $\Delta y$, the distribution of particles at later times 
will be given by the quantum mechanical distribution 
$\rho=\psi^*\psi$. In that sense, the quantum mechanical distribution 
at later times is achieved even in the limit $\Delta y\rightarrow 0$.
Therefore, there is no incompatibility
between Bohmian mechanics and standard quantum mechanics.

The source of the wrong conclusion of Akhavan and Golshani 
in \cite{ag} can also be understood in the following qualitative way. 
Assume that the width of the initial wavefunction is $\Delta y_i$,   
while that of the final wavefunction is $\Delta y_f$. Let their 
ratio be $R=\Delta y_f/\Delta y_i$. If the width of the initial 
distribution of particles is $\Delta y_i$, then the width 
of the final distribution of particles is equal to the 
product $\Delta y_i\cdot R$. The reasoning in \cite{ag} can be reduced 
to the following reasoning: one studies a case in which 
$\Delta y_i=0$, so one concludes that the final distribution 
has the width $\Delta y_i\cdot R=0\cdot R=0$. However, in this 
reasoning, one forgets that $R=\infty$. The correct reasoning is    
$\Delta y_i\cdot R=0\cdot \infty$, which is indeterminate. The 
regularization of the indeterminate expression leads to the result 
that the product is finite and equal to $\Delta y_f$, in agreement 
with standard quantum mechanics. 

Finally, note that a wavefunction proportional to a 
$\delta$-function is an idealization that never realizes in nature.
Therefore, even if the formal arguments in \cite{ag} were correct
(which, as argued above, were not), they would not correspond 
to an incompatibility of Bohmian mechanics with standard quantum 
mechanics in realistic situations.

This work was supported by the Ministry of Science and Technology of the
Republic of Croatia under Contract No. 0098002.

\end{document}